\documentclass[conference]{IEEEtran}

\IEEEoverridecommandlockouts
\usepackage{cite}
\usepackage{amsmath,amssymb,amsfonts}
\usepackage{algorithmic}
\usepackage{graphicx}
\usepackage{textcomp}
\usepackage{xcolor}
\usepackage{acronym}
\usepackage{bm}
\usepackage{hyperref}

\def\BibTeX{{\rm B\kern-.05em{\sc i\kern-.025em b}\kern-.08em
    T\kern-.1667em\lower.7ex\hbox{E}\kern-.125emX}}

\renewcommand{\baselinestretch}{0.92}

\acrodef{MC}{molecular communication}
\acrodef{pdf}{probability density function}
\acrodef{RX}{receiver}
\acrodef{TX}{transmitter}
\acrodef{CSI}{channel state information}
\acrodef{EM}{electro-magnetic}
\acrodef{DL}{deep learning}
\acrodef{AE}{autoencoder}
\acrodef{CSK}{concentration shift keying}
\acrodef{MOS}{metal oxide semiconductor}
\acrodef{SNR}{signal-to-noise ratio}
\acrodef{SER}{symbol error rate}
\acrodef{ODD}{odor delivery device}
\acrodef{RV}{random variable}
\acrodef{MDA}{mixture design algorithm}
\acrodef{AML}{approximate maximum likelihood}
\acrodef{SSER}{system SER}
\acrodef{GMoSK}{generalized molecule shift keying}

\newcommand{\x}{\mathbf{x}}

\newcommand{\xmax}{x_{\mathrm{max}}}

\newcommand{\xbar}{\mathbf{\bar{x}}}

\renewcommand{\H}{\mathbf{H}}

\newcommand{\y}{\mathbf{y}}
\newcommand{\ybar}{\bar{\mathbf{y}}}
\newcommand{\z}{\mathbf{z}}

\newcommand{\ntx}{\mathbf{n}^{\mathrm{TX}}}
\newcommand{\nc}{\mathbf{n}^{\mathrm{C}}}
\newcommand{\nrx}{\mathbf{n}^{\mathrm{RX}}}
\newcommand{\f}[1]{\mathbf{f}\left( #1 \right)}

\newcommand{\feasibleset}{\mathcal{F}_{\x}}

\newcommand{\nsymbols}{N}
\newcommand{\nspecies}{S}
\newcommand{\nsensors}{R}

\newcommand{\I}{\mathbf{I}}
\newcommand{\nullmatrix}{\mathbf{0}}

\newcommand{\zerovec}{\mathbf{0}}
\newcommand{\onevec}{\mathbf{1}}

\newcommand{\transpose}{^\mathrm{T}}

\newcommand{\muvec}{\boldsymbol{\mu}}

\newcommand{\covmat}{\mathbf{C}}

\newcommand{\nusers}{U}

\newcommand{\batchsize}{K}

\begin{document}

\title{Autoencoder-based Optimization of Multi-user Molecule Mixture Communication Systems}

\author{\IEEEauthorblockN{Bastian Heinlein$^{1,2}$, Nuria Zurita Jiménez$^{1}$, Kaikai Zhu$^{1}$, Sümeyye Carkit-Yilmaz$^{1}$,\\ Robert Schober$^{1}$, Vahid Jamali$^{2}$, and Maximilian Schäfer$^{1}$}
\IEEEauthorblockA{
\textit{$^1$ Friedrich-Alexander-Universität Erlangen-Nürnberg, Erlangen, Germany}\\
\textit{$^2$ Technical University of Darmstadt, Darmstadt, Germany}\\
\vspace*{-15mm}
}

}

\maketitle
\begin{abstract}
In this paper, we introduce an \ac{AE}-based scheme for end-to-end optimization of a multi-user molecule mixture communication system. In the proposed scheme, each transmitter leverages an encoder network that maps the user symbol to a molecule mixture. The mixtures then propagate through the channel to the receiver, which samples the channel using a non-linear, cross-reactive sensor array. A decoder network then estimates the symbol transmitted by each user based on the sensor observations. 
The proposed scheme achieves, for a given signal-to-noise ratio, lower symbol error rates than a baseline scheme from the literature in a single-user setting with full channel state information. We additionally demonstrate that the proposed \ac{AE}-based scheme allows reliable communication when the channel is unknown or changing. Finally, we show that for multiple access the system can account for different user priorities. 
In summary, the proposed \ac{AE}-based scheme enables end-to-end system optimization in complex scenarios unsuitable for analytical treatment and thereby brings \acl{MC} systems closer to real-world deployment.
\end{abstract}

\acresetall
\begin{figure*}[h]
    \centering
    \includegraphics[width=\textwidth]{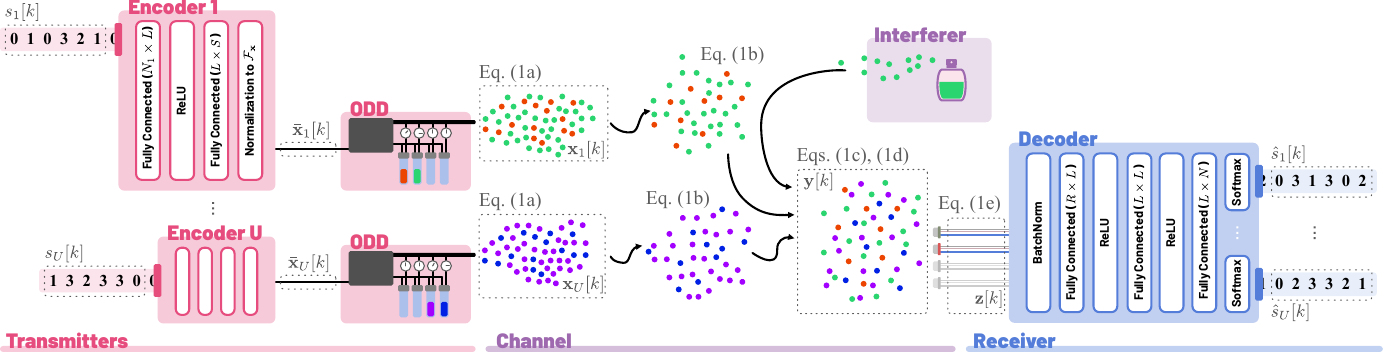}
    \vspace*{-7mm}
    \caption{\textbf{System Overview.} We consider a scenario where multiple TXs transmit information to a single RX. At the $i$-th TX, the corresponding encoder determines which mixture $\xbar_i[k]$ should be released for symbol $s[k]$ and sends the corresponding control signal to the \acl{ODD} (ODD). Due to hardware imperfections, the ODDs release mixtures $\x_i[k]$ which propagate through the channel, resulting in a molecule mixture $\y[k]$ at the RX. The decoder then leverages the output of the RX sensor array $\z[k]$ to compute symbol estimates $\hat{s}_1[k], \dots, \hat{s}_U[k]$.}
    \label{fig:system_overview}
\end{figure*}

\section{Introduction}\label{sec:introduction}
\Ac{MC} is an emerging research paradigm employing molecules for information transmission when \ac{EM} wave-based communication is not practical, e.g., for interfacing with biological entities or in \ac{EM}-wave denied environments~\cite{guo:molecular_physical_layer_6g}.
So far, the real-world deployment of \ac{MC} systems is hindered by the large gap between theoretical results, relying on unrealistic assumptions, and practical challenges such as incomplete \ac{CSI} or imperfect hardware~\cite{farsad:nn_detection_data_sequences_comm_systems,heinlein:nanocom}.

To address such challenges, \ac{DL} has emerged as a promising approach in recent years. For example, it has been shown in~\cite{farsad:nn_detection_data_sequences_comm_systems} that \ac{DL}-based detectors can outperform hand-crafted detectors when the channel conditions change or \ac{CSI} knowledge is limited. 
To further improve \ac{DL}-based detectors, \textbf{end-to-end optimization}, where transmission scheme and detector are jointly optimized, can be employed. Such end-to-end optimization usually relies on \acp{AE}, where an encoder maps the symbol, which is to be transmitted, to the desired physical signal. This signal is then sent over the channel so that the decoder can estimate the transmitted signal from noisy channel measurements. 

While this idea has been originally introduced in~\cite{oshea:introduction_dl_phy_layer} for \ac{EM}-based wireless communication, \ac{AE}-based end-to-end optimization has recently received considerable attention in the \ac{MC} literature~\cite{khanzadeh:end2endlearning_timevarying_higherorder_mc,khanzadeh:explainable_asymetric_ae_e2elearning_iobnt,mohamed:modelbased_e2e_mc_system_deepr_rl_ae,archetti:emergent_mc_preliminary_results}. However, most existing work has focused on nanoscale \ac{MC} - in fact, a recent review~\cite{gomez:communicating_smartly_mc_environments} identified no \ac{AE}-based approach for air-based \ac{MC}, where molecules propagate through the air over larger distances. 
Moreover, even on the nanoscale, the focus has been mostly on single-user settings with \ac{CSK}, i.e., a single molecule type is employed to transmit information~\cite{khanzadeh:end2endlearning_timevarying_higherorder_mc,khanzadeh:explainable_asymetric_ae_e2elearning_iobnt,mohamed:modelbased_e2e_mc_system_deepr_rl_ae}. 
This focus on \ac{CSK} ignores the opportunity for (i) higher data rates, (ii) multi-user communication, and (iii) robustness against unknown channel conditions by encoding information into \textbf{mixtures of molecules}~\cite{jamali:olfaction_inspired_MC,araz:rskm_time_varying_MC_channels}.
The only example of \ac{DL}-based mixture communication is~\cite{archetti:emergent_mc_preliminary_results}, where multiple users are positioned on a graph and learn a communication protocol encoding information in the concentrations $c_m$, $m \in \{1, \dots, \nspecies\}$, of $\nspecies$ molecule types. 
However, in~\cite{archetti:emergent_mc_preliminary_results} the \acp{RX} have direct access to the concentrations $c_m$ of each molecule type in the channel, only distorted by counting noise. 

In practice, however, the \ac{RX} can sample the channel only using imperfect sensors (e.g., an odorant receptor in natural or an \ac{MOS} sensor in synthetic systems). Such sensors are usually \textbf{non-linear and/or cross-reactive}. A sensor is non-linear if its response $r$ does not linearly scale with the concentration $c$ of a specific target molecule type $m$. 
A typical example of such non-linear behavior are power-laws that occur, e.g., in the sense of smell~\cite{teixeira:perception_fragrance_mixtures}, or in \ac{MOS} sensors~\cite{yamazoe:theory_power_laws_semiconducator_gas_sensors}. In this case, the sensor response follows $r \sim a \cdot c^b$, where $a,b \in \mathbb{R}$ are a scaling factor and the power-law coefficient, respectively. 
A sensor is cross-reactive if its response depends not only on a single species $m$, but on the concentration of multiple species. Examples of cross-reactive sensors are again natural odorant receptors or optical sensors, as molecules often have overlapping adsorption spectra~\cite{wietfeld:evaluation_multi_molecule_molecular_communication_testbed}. 
If a device is both non-linear and cross-reactive, its response $r$ is characterized by a non-linear function $r = f(c_1, \dots, c_\nspecies)$. Practical examples of such devices include \ac{MOS} sensors and odorant receptors in the saturation regime. 
These sensor characteristics are integral to the performance of communication systems since they determine which molecule concentrations can be distinguished by the \ac{RX} and which ones are easily confused. Yet, this topic has been mostly neglected in the \ac{MC} literature so far. Only recently, a scheme to optimize mixture alphabets for an \ac{RX} with a general non-linear, cross-reactive sensing array and a corresponding detector have been proposed in~\cite{heinlein:nanocom}. However, this work relies on perfect \ac{CSI} and considers only single-user systems. So far, unlike other fields such as optical fiber communications~\cite{karanov:e2e_dl_optical_fiber_communications,omidi:ae_optimized_pam_transceivers}, no end-to-end optimization scheme that accounts for the properties of the \ac{RX} hardware has been proposed for \ac{MC} systems.

In this work, we propose an \ac{AE}-based approach for end-to-end optimization of a general multi-user molecule mixture communication system with an \ac{RX} that employs a non-linear, cross-reactive sensing array. In particular, the main contributions of this work are as follows:
\begin{enumerate}
    \item We propose an \ac{AE}-based end-to-end optimization approach and show that for single-user systems it achieves lower \acp{SER} compared to the existing hardware-aware schemes from~\cite{heinlein:nanocom} that separately optimize the mixture alphabet and the detector.
    \item We show that our \ac{AE}-based approach is suitable for settings with unknown channel attenuation, thereby demonstrating its robustness to limited or changing \ac{CSI}.
    \item Finally, we extend the proposed \ac{AE}-based approach to a scenario where multiple \acp{TX} transmit symbols independently to a single \ac{RX}. We show that different importance can be assigned to the individual users, enabling calibration of the \acp{SER} of the \acp{TX}.
\end{enumerate}

\textbf{Notation:} Vectors and matrices are denoted by lowercase and uppercase bold letters, respectively. The $i$-th entry of vector $\x$ is denoted by $[\x]_i$ and the entry in row $i$ and column $j$ of matrix $\covmat$ as $[\covmat]_{ij}$. The transpose of matrix $\covmat$ is denoted by $\covmat\transpose$. 
$\I$ denotes the identity matrix, $\mathbf{1}$ denotes an all-one column vector, and $\nullmatrix$, depending on the context, either a column vector or a matrix containing only zeroes. 
Sets are shown by calligraphic letters and the cardinality of set $\mathcal{A}$ is denoted by $|\mathcal{A}|$. $\mathbb{R}$ and $\mathbb{R}_{\geq0}$ denote respectively the real and non-negative real numbers.

\section{System Model}\label{sec:system_model}
In this work, we consider a multi-user setting with $U$ independent \acp{TX}\footnote{In the following, we use the terms \textit{user} and \textit{\ac{TX}} interchangeably.}, as illustrated in Figure~\ref{fig:system_overview}. We assume that the $i$-th \ac{TX}, $i \in \{1, \dots, U\}$, aims to transmit one of $N_i$ symbols $s_i[k] \in \mathcal{S}_i = \{1, \dots, N_i\}$ to the \ac{RX} in the $k$-th symbol interval. Here, $N_i$ denotes the size of the symbol alphabet $\mathcal{S}_i$ of the $i$-th user, with $\sum_i N_i = N$. 
The $i$-th \ac{TX} maps symbol $s_i[k] \in \mathcal{S}_i$ to a mixture $\xbar_i[k] \in \mathcal{A}_i$ from its mixture alphabet $\mathcal{A}_i$, which is released by the \ac{ODD}. Here, $[\xbar_i[k]]_l$, $\forall l \in \{1, \dots, \nspecies \}$, denotes the expected concentration of the $l$-th molecule type close to the \ac{TX} after molecule release. 
In this work, we apply a per-molecule type constraint, i.e., we assume that $[\xbar_i[k]]_l\leq x_{\max}$, $\forall l \in \{1, \dots, \nspecies \}$, must be smaller than some maximum concentration $x_{\max}$ due to hardware constraints or safety considerations~\cite{hopper:multichannel_portable_odor_delivery_device}. Thus, each $\xbar_i[k]$ must lie in a feasible set $\feasibleset = [0, x_{\max}]^\nspecies$. 

Our system model is based on~\cite{heinlein:nanocom}, where a single-sample detector is employed to avoid the modeling of the intricate temporal channel dynamics and instead focus on the effects of the non-linear, cross-reactive \ac{RX} array. Moreover, the system model is  extended to a multi-user setting, leading to the following set of equations:
\begin{subequations}
    \begin{align}
        \x_i[k] &= \left[ \xbar_i[k] +\ntx_i[k] \right]_+ \;\;,\;\; i \in \{1, \dots, \nusers\} \label{eq:system_model:tx_noise} \\
        \ybar_i[k] &= \mathbf{H}_i \x_i[k]  \;\;,\;\; i \in \{1, \dots, \nusers\} \label{eq:system_model:propagation} \\
        \ybar[k] &= \sum_{i=1}^{\nusers} \ybar_i[k] \label{eq:system_model:accumulation} \\
        \y[k] &= \left[ \ybar[k] + \nc[k] \right]_+ \label{eq:system_model:channel_noise}\\
        \z[k] &= \f{\y[k]} + \nrx[k],\label{eq:system_model:rx_noise}
    \end{align}
\end{subequations}
where $[\cdot]_+ = \max\{0, \cdot\}$ is applied element-wise. 
Eq.~\eqref{eq:system_model:tx_noise} describes the \textbf{noisy release process of molecule mixtures} at each of the $\nusers$ users. Here, release noise $\ntx_i[k]$ reflects that \ac{TX} hardware like \acp{ODD} cannot perfectly control the amount of molecules released, resulting in a noisy released mixture $\x_i[k]$ at the $i$-th user. We model $\ntx_i[k]$ as normally distributed with mean $\muvec_{\ntx}$ and covariance matrix $\covmat_{\ntx}$. If the resulting $\x_i[k]$ would have negative entries due to $\ntx[k]$, we set those entries to zero.

Eqs.~\eqref{eq:system_model:propagation} and~\eqref{eq:system_model:accumulation} characterize the \textbf{expected concentration $\ybar[k]$ at the \ac{RX}}. Due to effects like the spatial spread of the molecules and/or atmospheric degradation, lower molecule concentrations $\ybar_i[k]$ due to the molecules released by the $i$-th \ac{TX} are expected at the \ac{RX} compared to $\x_i[k]$. The relationship between $\x_i[k]$ and $\ybar_i[k]$ is captured by diagonal channel matrix $\H_i$, where $[\H_i]_{jj}$ denotes the attenuation of the $j$-th molecule type released by the $i$-th user.
Assuming no interaction between the mixtures released by the different users, \eqref{eq:system_model:accumulation} then reflects the fact that the molecules released by the different \acp{TX} accumulate to $\ybar[k]$ at the \ac{RX}. 

Eq.~\eqref{eq:system_model:channel_noise} accounts for the impact of \textbf{propagation uncertainty and external interference}, affecting the \textit{actually} observed molecule concentrations at the \ac{RX}, $\y[k]$. In particular, we model the impact of these uncertainties by normally distributed channel noise $\nc[k]$ with mean $\muvec_{\nc}$ and covariance matrix $\covmat_{\nc}$ and set all negative entries in $\y[k]$ to zero to avoid negative concentrations. 

Finally,~\eqref{eq:system_model:rx_noise} relates the \textbf{noisy sensor response} $\z[k] \in \mathbb{R}^{\nsensors}$ to concentrations $\y[k]$, where $[\z[k]]_r$, $r \in \{1, \dots, \nsensors\}$, denotes the output of the $r$-th sensor and $R$ is the total number of sensors at the \ac{RX}. The mapping from $\y[k]$ to $\z[k]$ follows a non-linear function $\f\cdot$, which captures the non-linear, cross-reactive nature of the sensors. Note that we assume that the sensors are approximately memory-free, which is a realistic assumption if the molecule propagation dynamics are slow compared to the sensor dynamics. This is true for various types of practical sensors, including but not limited to optical sensors and many \ac{MOS} sensors~\cite{wietfeld:evaluation_multi_molecule_molecular_communication_testbed,dennler:high_speed_odor_sensing}. Additionally, the sensors introduce measurement noise $\nrx[k]$, e.g., thermal noise or quantization noise, modeled as normally distributed \ac{RV} with mean $\muvec_{\nrx}$ and covariance matrix $\covmat_{\nrx}$. %

\section{Auto-Encoder-based End-to-End Optimization}\label{sec:ae}
In this section, we describe the proposed \ac{AE} architecture and the loss function used for end-to-end system optimization.

\subsection{Auto-Encoder Architecture}
To enable end-to-end optimization, i.e., the joint optimization of the mixtures of each user and the detector, we employ an \ac{AE}-like architecture with one encoder per user and a single decoder at the \ac{RX} as depicted in Figure~\ref{fig:system_overview}. The encoder of each user deterministically\footnote{Due to its deterministic nature, the encoder can be replaced by a look-up table after training.} maps the symbol $s_i[k]$ to a mixture $\xbar_i[k]$. The encoder comprises two fully connected layers with $\mathrm{ReLu}$ activation functions after the first layer and sigmoid activation functions after the second layer. The first layer has $L=64$ neurons. The sigmoid activation functions are scaled so that the encoder output lies in $\feasibleset$. The output of each user's encoder, $\xbar_i[k]$, $i \in \{1, \dots, \nusers\}$, is then fed through the system model, i.e.,~\eqref{eq:system_model:tx_noise}-\eqref{eq:system_model:rx_noise}, resulting in sensor outputs $\z[k]$.
These are fed into the \ac{RX}'s decoder, which comprises a $\mathrm{BatchNorm}$ layer, followed by three fully connected layers, again with dimension $L=64$ in the latent space and $\mathrm{ReLu}$ activation functions. As shown in Figure~\ref{fig:system_overview}, the final layer's output of size $N$ is split into user-specific outputs of size $N_i$, $i \in \{1,\dots,\nusers\}$, each of which is followed by a $\mathrm{Softmax}$ activation function. The output for the $i$-th user is collected in vector $\hat{\mathbf{l}}_i$, where $[\hat{\mathbf{l}}_i]_s$ denotes the likelihood that symbol $s \in \{1, \dots, N_i\}$ has been transmitted. Due to its relatively small size and simple architecture, the decoder is suitable for deployment even on conventional micro-controllers.

\textit{\textbf{Remark:} Note that for end-to-end optimization via \acp{AE} that are trained using gradient descent, the system model must be differentiable to allow the flow of gradients to the encoders~\cite{archetti:emergent_mc_preliminary_results}. This condition can be fulfilled in all practical scenarios, because any $\f\cdot$ can be approximated by a neural network, which is per definition differentiable.}

\subsection{Loss Function}
Because each user transmits independently, we adopt a user-wise cross-entropy loss between $\hat{\mathbf{l}}_i$ and the true transmitted symbol~$s_i[k]$. However, some users may transmit more critical messages than others, e.g., when one user is responsible for security-critical monitoring while another one sends optional status updates. Therefore, we assign \textit{importance factor} $\lambda_i \geq 0$ to the $i$-th user, where a larger $\lambda_i$ implies that the symbols of the $i$-th user are more important.
This results in the following loss function:
\vspace*{-2mm}
\begin{equation}\label{eq:loss}
    \mathcal{L} = \sum_{i=1}^{\nusers} \lambda_i \sum_{k=1}^{\batchsize} \mathrm{CrossEntropy}\left(\hat{\mathbf{l}}_i , s_i[k]\right),
\end{equation}
where $K$ and $\mathrm{CrossEntropy}\left(\cdot, \cdot\right)$ denote respectively the batch size and the cross-entropy loss~\cite{oshea:introduction_dl_phy_layer}. 

\section{Evaluation}\label{sec:eval}
\subsection{Simulation Setup}
In our simulations, we set $\xmax=2 \cdot 10^4$ (in parts per million) for each of the $\nspecies=3$ molecule types and all entries $[\H_i]_{jj} = h_i=h = 0.01$ for simplicity. 
The \ac{RX} employs $\nsensors=2$ \ac{MOS} sensors whose non-linear response follows a power-law behavior. Because there is not sufficient data in the literature that describes the response of commercially available \ac{MOS} sensors to molecule mixtures with more than two molecule types, we follow the approach described in \cite[Appendix~A]{heinlein:journal} to generate artificial \ac{MOS} sensors. 
We parametrize \eqref{eq:system_model:tx_noise}-\eqref{eq:system_model:rx_noise} as follows: The noise sources have mean values $\muvec_{\ntx}=\zerovec$, $\muvec_{\nc}=10 \cdot \onevec$, $\muvec_{\nrx}=\zerovec$ and covariance matrices $\covmat_{\ntx}= \nu \cdot 10^6 \cdot \I$, $\covmat_{\nc}= \nu \cdot 10 \cdot \I$, $\covmat_{\nrx}= \nu \cdot 10^{-13} \cdot \I$. 
Here, $\nu$ scales the power of all noise sources, controlling the \ac{SNR}\footnote{Note that there is no analytical expression for the \ac{SNR}, because \ac{TX} and channel noise are fed through non-linear functions.}. 

We train the encoders and decoders jointly using gradient-descent by minimizing~\eqref{eq:loss}. To this end, we employ the $\mathrm{Adam}$ optimizer with learning rate $10^{-3}$ over 250 epochs with 5 batches of size $K = 256$ symbols each~\cite{oshea:introduction_dl_phy_layer}. To avoid training a new \ac{AE} for each \ac{SNR}, we choose for each batch a random noise level $\nu = 10^{\frac{l}{5}}$, $l \in \{-5, -4, \dots, 5\}$, in all scenarios.

\subsection{Static Point-to-Point Link with Full CSI}
\begin{figure}
    \centering
    \includegraphics[width=\linewidth]{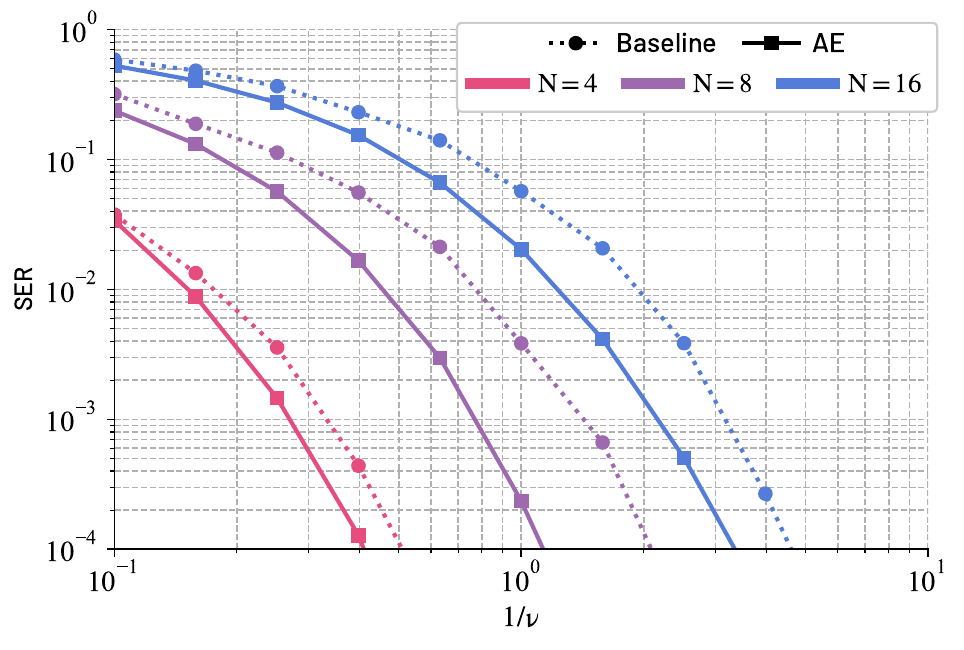}
    \vspace*{-10mm}
    \caption{\textbf{Full \ac{CSI}.} We show the \ac{SER} of the proposed \ac{AE}-based scheme (solid) and the baseline scheme from~\cite{heinlein:nanocom} (dotted) as a function of $1/\nu$ for different alphabet sizes $N$ (indicated by different colors).}
    \label{fig:eval:point_to_point_full_CSI}
\end{figure}
\begin{figure*}[t!]
    \centering
    \includegraphics[width=\linewidth]{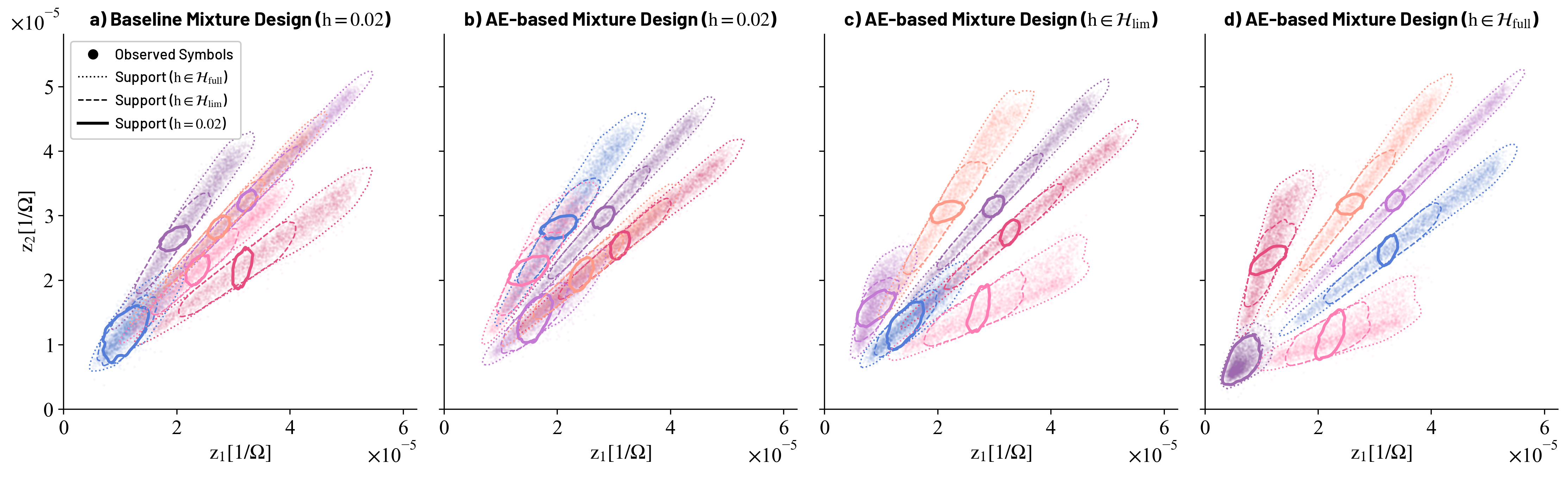}
    \vspace*{-8mm}
    \caption{\textbf{Sensor Outputs for Unknown Channel Attenuation.} We show the sensor outputs (dots) for the different symbols (indicated by different colors). The solid, dashed, and dotted lines indicate respectively the areas where $95\%$ of the symbols for $h=0.02$, $h \in \mathcal{H}_{\mathrm{lim}}$, and $h \in \mathcal{H}_{\mathrm{full}}$ lie.}
    \label{fig:eval:unknown_channel_gain_mixtures}
\end{figure*}
First, we evaluate the proposed \ac{AE}-based scheme in a single-user scenario, i.e., $\nusers=1$, with full \ac{CSI}, i.e., fully known system parameters. 
We compare the proposed scheme to the methods proposed in~\cite{heinlein:nanocom}, where a heuristic \ac{MDA} and an \ac{AML} detector have been proposed: The \ac{MDA} iteratively builds a mixture alphabet by selecting mixtures whose expected sensor outputs are well-separable from the other mixtures in the alphabet. The \ac{AML} detector approximates the sensor output distribution of each symbol for a given alphabet as a normal distribution, which is parametrized by first- and second-order moments which account for all noise sources in the system and the non-linear sensor behavior. 

In Figure~\ref{fig:eval:point_to_point_full_CSI}, we show the resulting \acp{SER} of the baseline scheme (dotted) and the proposed \ac{AE}-based scheme for alphabet sizes $\nsymbols \in \{4, 8, 16\}$. For all alphabet sizes and values of $1/\nu$, the proposed scheme achieves a significantly lower \ac{SER} compared to the baseline algorithm, demonstrating the superiority of \ac{AE}-based end-to-end optimization.

\subsection{Unknown Channel Attenuation}
\begin{figure}
    \centering
    \includegraphics[width=\linewidth]{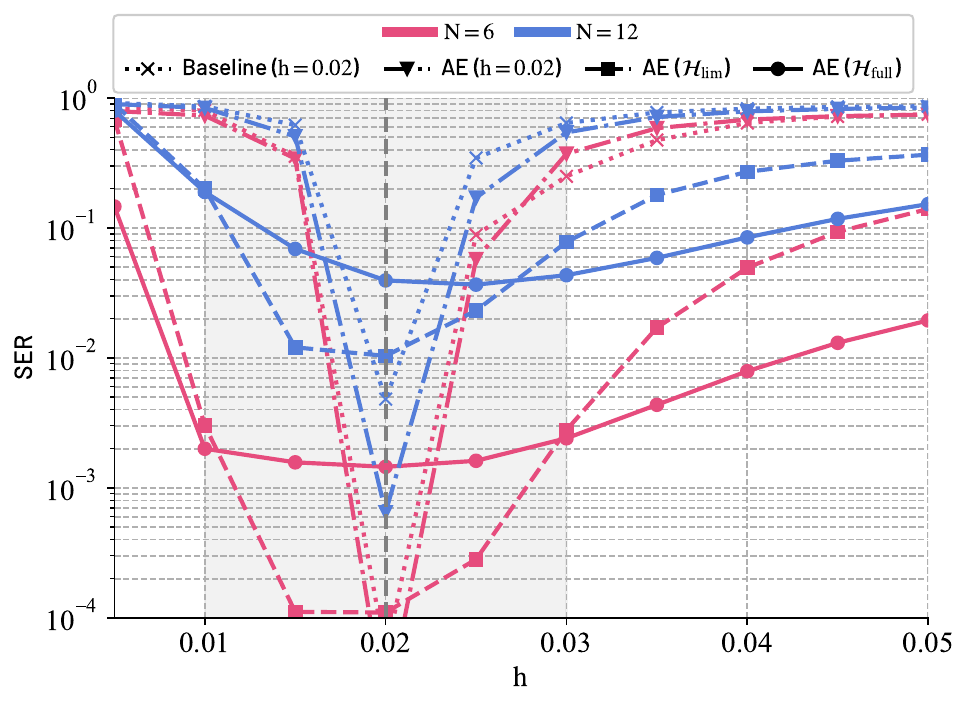}
    \vspace*{-9mm}
    \caption{\textbf{\acp{SER} for different channel attenuations.} We show the \acp{SER} of four different schemes as a function of the channel attenuation $h$ and different alphabet sizes $\nsymbols$.}
    \label{fig:eval:unknown_channel_gain_ser}
    \vspace{-5mm}
\end{figure}
Next, we consider a scenario where only limited \ac{CSI} is available. In particular, we assume that the channel attenuation $h$ can take any value in interval $\mathcal{H}_{\mathrm{full}} = [h_{\min}, h_{\max}]$.
To account for this, we adjust the \ac{AE} training by drawing $h$ randomly from $\mathcal{H}_{\mathrm{full}}$ for each symbol. 

Figure~\ref{fig:eval:unknown_channel_gain_mixtures} shows the resulting sensor outputs of four different schemes in a single-user setting with $N=|\mathcal{A}|=6$, $h_{\min} = 0.005$, $h_{\max} = 0.05$, and $\nu=1$. 
The solid, dashed, and dotted lines indicate respectively the areas where $95\%$ of the symbols for $h=0.02$, $h \in \mathcal{H}_{\mathrm{lim}}$, and $h \in \mathcal{H}_{\mathrm{full}}$ lie, where $\mathcal{H}_{\mathrm{lim}} = [0.01, 0.03]$. Different colors indicate different symbols. 
In Figure~\ref{fig:eval:unknown_channel_gain_mixtures}-a), we show the sensor outputs for the baseline scheme from~\cite{heinlein:nanocom} which optimizes the mixture alphabet for $h=0.02$. 
Clearly, the outputs corresponding to $h=0.02$ are well-separable. However, the outputs $\z[k]$ of different symbols for $\mathcal{H}_{\mathrm{lim}}$ and $\mathcal{H}_{\mathrm{full}}$ are overlapping. 
In Figure~\ref{fig:eval:unknown_channel_gain_mixtures}-b), we show the sensor outputs when an \ac{AE} is trained on $h=0.02$, showing the same qualitative behavior as the baseline scheme. 
In Figure~\ref{fig:eval:unknown_channel_gain_mixtures}-c), we show $\z[k]$ for an \ac{AE} trained on $\mathcal{H}_{\mathrm{lim}}$. In contrast to schemes in Figures~\ref{fig:eval:unknown_channel_gain_mixtures}-a) and~\ref{fig:eval:unknown_channel_gain_mixtures}-b), this scheme found an alphabet resulting in separable sensor outputs for $\mathcal{H}_{\mathrm{lim}}$. However, the outputs for $\mathcal{H}_{\mathrm{full}}$ are still significantly overlapping.
Finally, in Figure~\ref{fig:eval:unknown_channel_gain_mixtures}-d), we show the sensor outputs belonging to the \ac{AE} trained on $\mathcal{H}_{\mathrm{full}}$. For this \ac{AE}, the sensor outputs are almost perfectly separable for the full range of $h$. %

For a more quantitative evaluation, we show in Figure~\ref{fig:eval:unknown_channel_gain_ser} the \acp{SER} of all four schemes as a function of $h$ and for $N \in \{6, 12\}$. When comparing the different schemes for a given $N$, several observations can be made: The baseline (dotted line) and the $h$-specific \ac{AE} (dash-dotted line) achieve the lowest \acp{SER} for $h=0.02$. However, their \acp{SER} increase dramatically even for small mismatches of $h$. 
In contrast, the \ac{AE} trained on $\mathcal{H}_{\mathrm{full}}$ (solid line) achieves, averaged over all $h$, the lowest \acp{SER}. At the same time, it has the highest \acp{SER} for $h=0.02$. Interestingly, the \acp{SER} of this scheme are not identical for all $h$: As $h$ becomes very small, the \ac{SER} increases because only very few molecules reach the \ac{RX} at all. Similarly, the \ac{SER} also increases as $h$ becomes large. This can be explained based on Figure~\ref{fig:eval:unknown_channel_gain_mixtures}-d): The sensor outputs belonging to the symbols indicated by purple and red colors overlap when $h$ is low for red or $h$ is high for purple. Because the \ac{RX} does not have \ac{CSI} knowledge, it cannot distinguish both scenarios, resulting in a high \ac{SER}.
Finally, the \ac{AE} trained on $\mathcal{H}_{\mathrm{lim}}$ (dashed line) achieves a trade-off between $h$-specific training and training over the full range of $h$. Interestingly, it generalizes beyond its trained range, resulting in lower \acp{SER} then the $h$-specific schemes.

\subsection{Multi-User Scenario}
\begin{figure}
    \vspace*{-8mm}
    \centering
    \includegraphics[width=\linewidth]{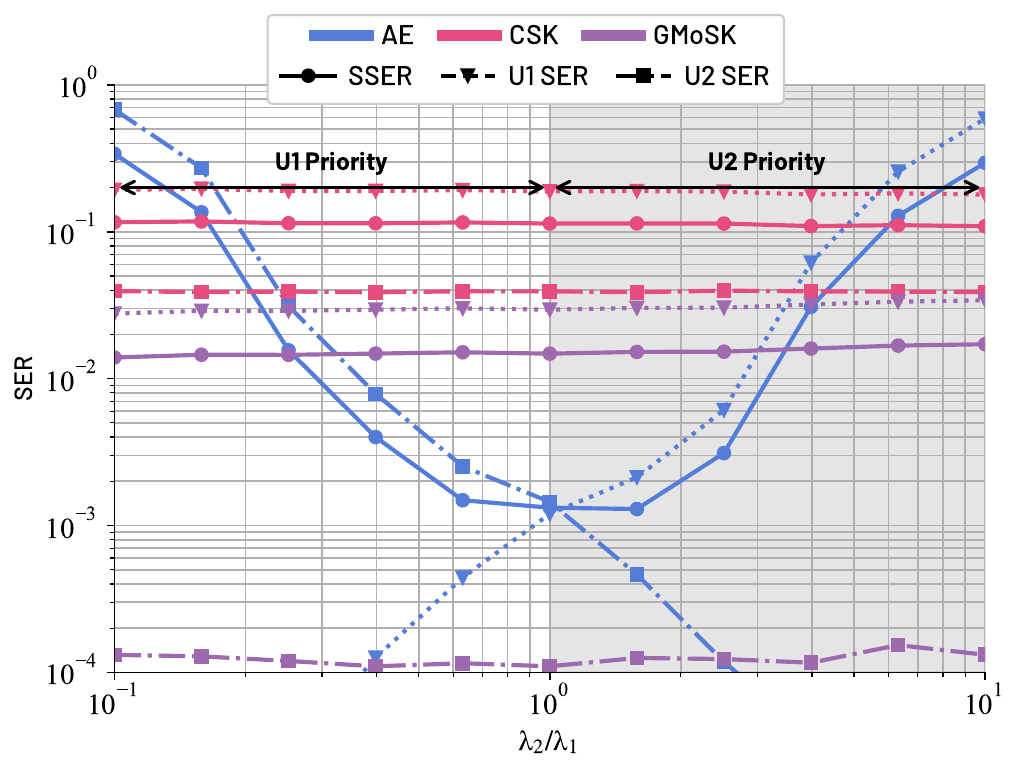}
    \vspace*{-8mm}
    \caption{\textbf{User Importance.} We compare the user-specific \acp{SER} and the \acl{SSER} (SSER) for \ac{AE}-based end-to-end optimization and \ac{CSK} modulation of each user for various importance factors.}
    \label{fig:eval:multi_user}
    \vspace{-5mm}
\end{figure}
Finally, we consider a multi-user setting where two \acp{TX} send messages independent of each other to a single \ac{RX}. Here, we assume a symmetric but known channel with $h_1 = h_2=10^{-2}$, $\xmax=1.5 \cdot 10^4$, $\nspecies=4$, $\nsensors=3$, and $\nsymbols_1 = \nsymbols_2 = 4$. 
As baseline, we assign two of the four molecule types to each user. Then, each user employs a conventional modulation scheme using its assigned molecule types. We consider both \ac{CSK} and \ac{GMoSK}~\cite{chen:gmosk} as modulation schemes and train a decoder network at the \ac{RX} for detection. For \ac{CSK}, each user employs one of its molecule types, i.e., $[\xbar[k]]_{l_{i,1}} \in \{0, \frac{1}{3}\xmax, \frac{2}{3} \xmax, \xmax \}$, where $l_{i,1}$ is the index of the first molecule type assigned to user $i$. 
For \ac{GMoSK}, each user employs both of its assigned molecules, resulting in $[\xbar[k]]_{l_{i,1}} \in \{0, 0, \xmax, \xmax \}$ and $[\xbar[k]]_{l_{i,2}} \in \{0, \xmax, 0, \xmax\}$, where $l_{i,2}$ is the index of the second molecule type assigned to user $i$. 

In Figure~\ref{fig:eval:multi_user}, we show the \ac{SER} of each user and the \ac{SSER}, i.e., the average \ac{SER}, as a function of the relative importance of both users, represented by the ratio $\lambda_2/\lambda_1$ of their importance factors, for $\nu=1$. 
Note that we report the results of the baseline schemes only for the molecule assignment that yields minimum \ac{SSER} for $\lambda_2/\lambda_1=1$. 

First, we observe that for $\lambda_2/\lambda_1=1$, i.e., identical user priorities, the \ac{AE}-based scheme achieves the lowest \ac{SSER} and both users have identical \acp{SER}. In contrast, both \ac{CSK} and \ac{GMoSK} yield higher \acp{SSER} and both users have different \acp{SER}. This is due to the fact that the sensor array of the \ac{RX} is not equally sensitive to all molecule types. Therefore, one user will typically have a lower \ac{SER} than the other. 

Second, we observe that both baselines are unaffected by varying $\lambda_2/\lambda_1$. This is expected as the mixture alphabets, and not the detector, determines the overlap of the sensor outputs for different symbols, which governs the achievable \acp{SER}. %
In contrast to the baselines, the \ac{AE}-based approach can design the mixture alphabets to achieve different \acp{SER} for different users. As one user becomes favored, its \ac{SER} decreases at the expense of a higher \ac{SER} for the other user. This demonstrates that the proposed \ac{AE}-based end-to-end optimization does not only outperform heuristically designed mixture alphabets in multi-user settings but also enables the consideration of different priorities for different users.

\section{Conclusion and Future Work}\label{sec:conclusion}
In this work, we studied the \ac{AE}-based end-to-end optimization of multi-user molecule mixture communication systems with non-linear, cross-reactive sensor arrays at the \ac{RX}. Specifically, we showed that the end-to-end-optimized mixture communication system achieves a lower \ac{SER} compared to a baseline scheme from the literature even in static single-user settings with perfect \ac{CSI}. 
In more challenging settings, where the channel attenuation is unknown, the \ac{AE}-based system finds mixtures that are well-separable for a wide range of channel attenuations while baseline schemes do not generalize to unknown attenuations. 
Finally, we showed that the \ac{AE}-based end-to-end optimization enables reliable communication even if multiple users transmit mixtures independently over a shared channel. 
Interesting directions for future work include the incorporation of signal-dependent noise, accounting for sensor failures, and extending the proposed system to multiple \acp{RX}.

\vspace*{-3mm}
\section*{Acknowledgements}
This work was funded by the Deutsche Forschungsgemeinschaft (DFG, German Research Foundation) – GRK 2950 – Project-ID 509922606. This project has received funding from the European Union’s Horizon Europe – HORIZON-EIC-2024-PATHFINDEROPEN-01 under grant agreement Project N. 101185661.

\vspace*{-3mm}
\bibliographystyle{ieeetr}
\bibliography{literature}
\clearpage
\end{document}